\documentclass{article}

    \usepackage[preprint,nonatbib]{neurips_2020}

\usepackage[utf8]{inputenc} 
\usepackage[T1]{fontenc}    
\usepackage{hyperref}       
\usepackage{url}            
\usepackage{booktabs}       
\usepackage{siunitx}
\usepackage{amsfonts}       
\usepackage{nicefrac}       
\usepackage{microtype}      
\usepackage{diagbox}
\usepackage{wrapfig}
\usepackage{multirow}
\usepackage{graphicx}
\usepackage{float}
\usepackage{wrapfig}
\usepackage{enumitem}
\usepackage{subfigure}
\title{XRayGAN: Consistency-preserving Generation of X-ray Images from Radiology Reports}

\author{%
  Xingyi Yang \\
  University of California San Diego\\
  \texttt{x3yang@eng.ucsd.edu}
  \And
  Nandiraju Gireesh \\
  Birla Institute of Technology and Science, Pilani\\
  \texttt{f20170720@hyderabad.bits-pilani.ac.in}\\
  \And
  Eric Xing\\
  Carnegie Mellon University\\
  \texttt{epxing@cs.cmu.edu}
  \And
  Pengtao Xie \\
  University of California San Diego\\
  \texttt{pengtaoxie2008@gmail.com}
}

\begin{document}

\maketitle

\begin{abstract}
To effectively train medical students to become qualified radiologists, a large number of X-ray images collected from patients with diverse medical conditions are needed. However, due to data privacy concerns, such images are typically difficult to obtain. To address this problem, we develop methods to generate view-consistent, high-fidelity, and high-resolution X-ray images from radiology reports to facilitate radiology training of medical students. This task is presented with several challenges. First, from a single report, images with different views (e.g., frontal, lateral) need to be generated. How to ensure consistency of these images (i.e., make sure they are about the same patient)? Second, X-ray images are required to have high resolution. Otherwise, many details of diseases would be lost. How to generate high-resolutions images? Third, radiology reports are long and have complicated structure. How to effectively understand their semantics to generate high-fidelity images that accurately reflect the contents of the reports? To address these three challenges, we propose an XRayGAN composed of three modules: (1) a view consistency network that maximizes the consistency between generated frontal-view  and lateral-view images; (2) a multi-scale conditional GAN that progressively generates a cascade of images with increasing resolution; (3) a hierarchical attentional encoder that learns the latent semantics of a radiology report by capturing its hierarchical linguistic structure and various levels of clinical importance of words and sentences. Experiments on two radiology datasets demonstrate the effectiveness of our methods. To our best knowledge, this work represents the first one generating consistent and high-resolution X-ray images from radiology reports. The code is available at \url{https://github.com/UCSD-AI4H/XRayGAN}.

\end{abstract}
\vspace{-0.2cm}
\section{Introduction}
\vspace{-0.3cm}

\begin{figure}[t]
    \centering
    \includegraphics[width=0.8\linewidth]{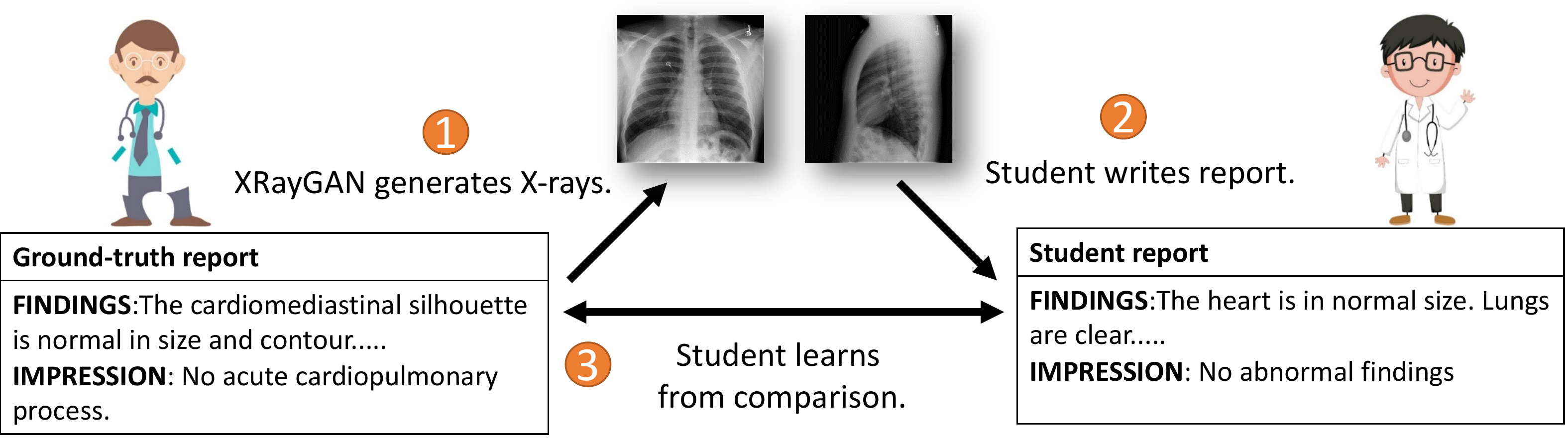}
    \caption{An illustration on how XRayGAN can be used for training radiology students. 
    }

    \label{fig:student training}
    \vspace{-4mm}
\end{figure}

X-ray is a widely-used imaging technique for making diagnostic and treatment decisions. Medical professionals interpreting X-ray images are called radiologists. To train medical students into radiologists, a lot of X-ray images are needed, taken from patients with various medical conditions, medical histories, demographics, etc. However, due to data privacy concerns, such images are typically difficult to access. As a result, medical students are trained using a limited number of de-identified X-rays which are not rich enough to serve the purpose of training. To address this problem, we aim to develop computer vision methods to automatically generate  clinically-diverse,  high-resolution, and high-fidelity X-rays  which provide a rich resource for medical student training. 
Figure \ref{fig:student training} illustrates the training process. A medical educator writes down a radiology report describing the lesions (e.g., pneumonia, effusion) of a patient and the attributes of the lesions such as locations, severity, appearance, size, etc. Our tool takes this report as input and automatically generates X-ray images that reflect the lesions and their attributes described in the report. Then the medical students are asked to read these generated X-ray images and write down radiology reports to narrate the clinical findings in the images. The reports written by students are compared with the ground-truth report written by the educator. Students can learn from the comparisons.

\begin{wrapfigure}{r}{0.6\textwidth}
     \centering
     \vspace{-4.3mm}
     \includegraphics[width=0.6\columnwidth]{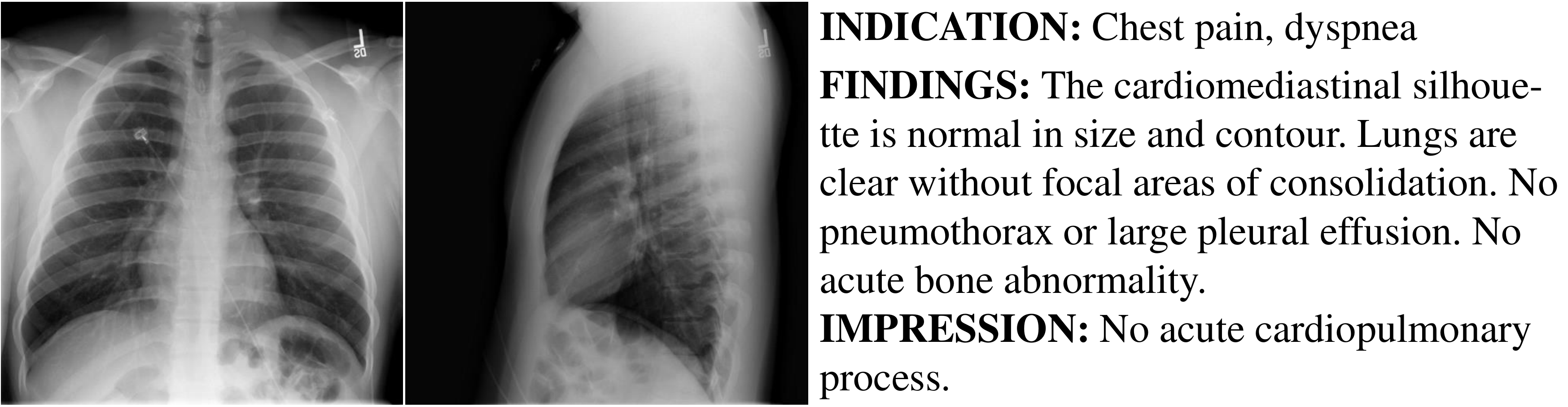}
     \vspace{-5mm}
     \caption{(Left) A frontal-view X-ray image. (Middle) A lateral-view image. These two images are from the same patient. (Right) A radiology report written from these two images. Each report contains  multiple sentences.}
     \label{fig:sample}
     \vspace{-3mm}
 \end{wrapfigure}

While this is a promising solution for mitigating the deficiency of radiology training resources, generating high-fidelity X-ray images from radiology reports is highly challenging. First, in clinical practice, a radiology report is typically written by interpreting X-ray images taken from different views, such as frontal view, lateral view, etc., as shown in Figure \ref{fig:sample}. When generating images from a radiology report, instead of generating a single image, we need to generate multiple images corresponding to different views. And these images from different views need to be clinically consistent, i.e., they reflect the medical conditions of the same patient. Second, X-ray images are usually of high-resolution so that radiologists could identify subtle clinical findings. How to generate high-resolution images is technically demanding. Third, the radiology reports are long and have complicated syntax and semantics. The reports have  hierarchical structures: each report has multiple sentences; each sentence has a sequence of words. Different words and sentences have different levels of clinical importance. How can we generate high-fidelity images that accurately reflect the complicated semantics in the reports? In this work, we aim to address these challenges. To our best knowledge, our work represents the first one that generates multi-view high-resolution medical images from clinical texts. 

To generate view-consistent images, we train a view consistency network (VCN) which takes a frontal-view image and a lateral-view image as inputs and judges whether they belong to the same patient. This network is trained offline using real X-ray images where positive examples are pairs of frontal-view and lateral-view images belonging to the same patient and negative examples are those from different patients. When training the X-ray generation models, we use the pre-trained VCN to evaluate how consistent the generated images are and encourage the generators to generate images with high consistency. To generate high-resolution X-ray images, we adopt a multi-scale conditional progressive generation approach~\cite{karras2018progressive}. Considering that it is relatively easy to generate high-fidelity images when the size of images is small, we first generate small-sized low-resolution images using a conditional GAN~\cite{Mirza2014ConditionalGA} which is conditioned on the latent embedding of the input report. Then starting from the high-fidelity but low-resolution image, we generate images with increasing resolutions progressively in multiple stages. To effectively understand the semantics of radiology reports, we develop a hierarchical attentional encoder, which uses a word-level LSTM~\cite{hochreiter1997long} network to capture the local semantics within a sentence and a sentence-level LSTM to capture the long-range semantics across sentences. Attentional modules~\cite{vaswani2017attention} are used to identify words and sentences that are clinically more important. On two radiology datasets, we demonstrate the effectiveness of our methods in generating view-consistent, high resolution, and high-fidelity X-rays from radiology reports. 

The major contributions of this paper are as follows:
\begin{itemize}[leftmargin=*]
    \item We propose an XRayGAN to generate view-consistent, high-fidelity, and high resolution X-ray images from radiology reports. To our best knowledge, our work is the first of its kind. 
    \item We propose a view consistency network to encourage the generated frontal-view image and lateral-view image to be consistent (i.e., coming from the same patient).
    \item We develop a multi-scale conditional GAN to generate high-resolution X-ray images.
    \item We develop a hierarchical attentional encoder to effectively understand the semantics of long and complicated radiology reports.
    \item We demonstrate the effectiveness of our model on two radiology datasets.
\end{itemize}

The rest of the paper is organized as follows.  Section 2 introduces the method. Section 3 presents the experimental results. Section 4 reviews related works. Section 5 concludes the paper. 

\section{Method}
\vspace{-0.2cm}

\begin{figure}[t]
    \centering
    \includegraphics[width=\linewidth]{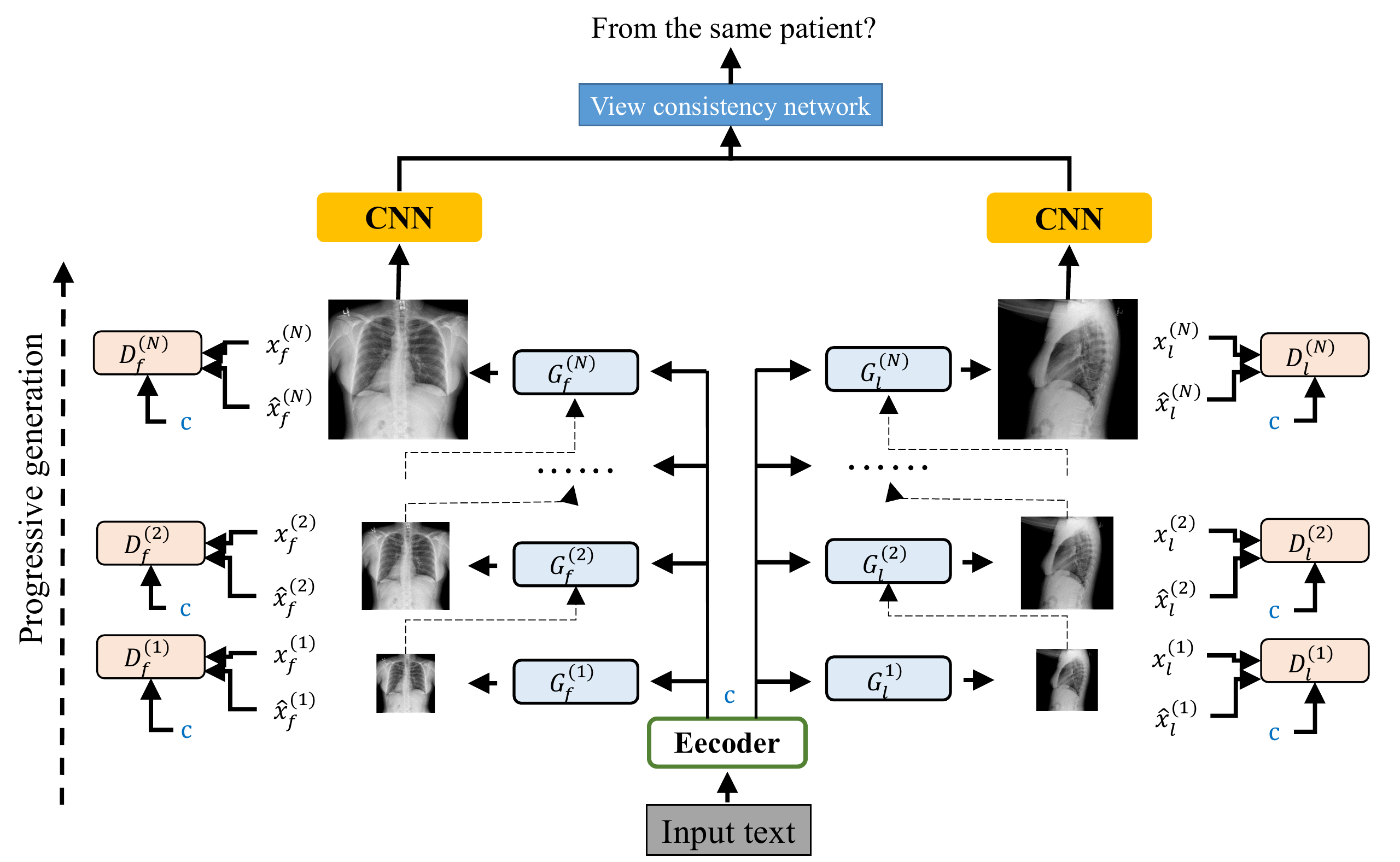}
    \vspace{-4mm}
    \caption{Overview of XRayGAN. Our method generates realistic bi-planer chest radiography in a coarse-to-fine manner, which consists of a hierarchical attentional encoder, a multi-scale conditional GAN, and a view consistency network.}
    \label{fig:pipeline}
    \vspace{-3mm}
\end{figure}

We aim to develop a method which takes a radiology report as input and generates a set of X-ray images that manifest the clinical findings narrated in the report. For simplicity, we assume there are two views of images: frontal view and lateral view, each with one image. It is straightforward to extend our method to more views (each with more than one image). 

At a high level, our model consists of three modules: hierarchical attentional encoder (HAE),  multi-scale conditional GAN (MSCGAN), and view consistency network (VCN). The HAE takes the radiology report as input and generates an encoding vector that captures the semantics of the report. The MSCGAN takes the report encoding generated by HAE as input and generates a frontal-view image and a lateral-view image. The VCN takes the two images generated by MSCGAN as inputs and produces a score that measures how consistent the frontal-view image is with the lateral-view image. Figure \ref{fig:pipeline} shows the overview of our model.  In the following sections, we describe each module in detail.

\subsection{Hierarchical Attentional Encoder}
\label{section:HAE}

The HAE takes the radiology report as input and encodes it into an embedding vector. The report has a hierarchical linguistic structure: it  is composed of a sequence of sentences and each sentence is composed of a sequence of words. To capture such a hierarchical structure, we use a hierarchical LSTM network~\cite{yang2016hierarchical}, which consists of two levels of LSTM networks: word-level LSTM and sentence-level LSTM. The word-level LSTM takes the sequence of words in one sentence as inputs and learns  latent embeddings for each word and the sentence. The sentence-level LSTM takes the embeddings of a sequence of sentences as inputs and learns an embedding of the report. Within a sentence, some words (e.g., medical terms) are more important than others (e.g., functional words). To account for the different importance of different words, we use a word-level attentional~\cite{yang2016hierarchical} module which calculates an attention score for each word and these attention scores are used to re-weight the word embeddings when generating the embedding of the entire sentence. Similarly, at the sentence level, certain sentences are more important than others. A sentence-level attention mechanism is used to account for this fact. 

\textbf{Attentional Word-level LSTM} Given a sequence of $T$ words in a sentence, we use an attentional word-level LSTM to learn a representation for this sentence. For a word $w^{(t)}$ (represented using one hot encoding) at position $t$, we obtain its embedding $\mathbf{c}^{(t)}_{word} = \mathbf{W}_ew^{(t)}$ by querying a learnable embedding matrix $\mathbf{W}_e$. Then  $\mathbf{c}^{(t)}_{word}$ is fed into a bi-directional LSTM~\cite{graves2005framewise} which produces hidden states $\overrightarrow{\mathbf{h}}_{word}^{(t)}$  and $\overleftarrow{\mathbf{h}}_{word}^{(t)}$ that capture contextual information of $w^{(t)}$.
Different words have different importance. We use an attentional module~\cite{attention} to learn the importance of each word. The importance $e_{word}^{(t)}$ of $w^{(t)}$ is calculated as: $e_{word}^{(t)}= \mathbf{v}_w\tanh(\mathbf{W}_{w}\mathbf{h}_{word}^{(t)}+\mathbf{b}_{w})$, 
where $\mathbf{h}_{word}^{(t)}$ is the concatenation of $\overrightarrow{\mathbf{h}}_{word}^{(t)}$  and $\overleftarrow{\mathbf{h}}_{word}^{(t)}$, $\mathbf{W}_{w}$ is a learnable weight matrix, $\mathbf{v}_w$ and $\mathbf{b}_{w}$ are learnable weight vectors. Then we normalize these importance scores using softmax: $\alpha_{word}^{(t)}= \exp(e_{word}^{(t)})/\sum_{j=1}^T \exp(e_{word}^{(j)})$. Finally, the representation of this sentence is calculated as a weighted sum of words' representations: $\mathbf{c}_{sent}= \sum_{j=1}^T \alpha_{word}^{(j)}\mathbf{h}_{word}^{(j)}$, where more important words contribute more in representing the sentence.

\textbf{Attentional Sentence-level LSTM} Given a sequence of sentences in the report, we first use the word-level LSTM to obtain an embedding of each sentence. Then we feed these embeddings into a sentence-level LSTM to obtain an embedding of the  report. Similar to the word-level LSTM, the sentence embeddings are fed into a bi-directional LSTM to capture the contextual information. An attentional module is used to calculate the importance of each sentence. The normalized importance scores are used to take a weighted sum of the hidden states produced by the bi-directional LSTM, yielding a representation of the report.

\vspace{-0.1cm}
\subsection{Multi-scale conditional GAN (MSCGAN)}
\vspace{-0.1cm}
The MSCGAN takes the embedding of the report as input and generates a frontal-view image and a  lateral-view image. X-ray images have a high requirement of resolution. Many details of  lesions would be lost if the resolution is low. To generate high resolution images, MSCGAN adopts a multi-scale progressive generation approach~\cite{Shaham_2019_ICCV}: first generate high-quality but low-resolution images, then generate higher-resolution images from lower-resolution ones progressively in multiple stages. 
The MSCGAN consists of two types of generators: (1) basic generators which generate frontal-view and lateral-view images with the lowest resolution; (2) progressive generators which take a generated lower-resolution image as input and generate a higher-resolution image. These generators are all conditional generators, where the generation of images is conditioned on the embedding of the radiology report.  

\textbf{Basic generators} The basic generators $G_f^{(1)}$ and $G_l^{(1)}$ take the embedding $\mathbf{c}$ of the radiology report as input and generate a frontal-view X-ray $x_f^{(1)}$ and a lateral-view X-ray $x_l^{(1)}$ with the lowest resolution: $x_f^{(1)} = G_f^{(1)}(\mathbf{c})$, $x_l^{(1)} = G_l^{(1)}(\mathbf{c})$.  Since the resolution of $x_f^{(1)}$ and $x_l^{(1)}$ is low, it is relatively easy to generate them. The basic generators capture coarse-grained information in images, determining how the X-rays look roughly. The fine-grained details will be filled in by the progressive generators.

\textbf{Progressive generators} Given the images generated by basic generators, a cascade of progressive generators (PGs) produces a sequence of images with increasing resolutions. At each stage in the cascade, the image resolution increases by a factor of 2. The outputs of PGs at stage $n-1$  are the inputs of PGs at stage $n$. PGs at all stages are conditioned on the embedding $\mathbf{c}$ of the report. Given the frontal-view image $x_f^{(n-1)}$ and lateral-view image $x_l^{(n-1)}$ generated at stage $n-1$,  PGs $G_f^{(n)}$ and $G_l^{(n)}$ at stage $n$ generate two higher-resolution images containing more fine-grained visual information:
\begin{equation}
    x_f^{(n)} = \alpha G_f^{(n)}(x_f^{(n-1)},\mathbf{c}) + (1-\alpha)U(x_f^{(n-1)}),\quad  x_l^{(n)} = \alpha G_l^{(n)}(x_l^{(n-1)},\mathbf{c}) + (1-\alpha)U(x_l^{(n-1)})
\end{equation}
where $U(\cdot)$ is an operation that resizes $x^{(n-1)}$ to the size of $x^{(n)}$. The higher-resolution image is a linear combination of the generated image and the resized lower-resolution image. $\alpha=0.5$ controls the level of blending.

\textbf{Discriminators} Similar to GAN~\cite{Shaham_2019_ICCV}, we use discriminators at each stage to judge whether images are generated or real. The generators are learned in a way that such a discrimination is difficult to achieve. The discriminators are based on patches whose sizes are different at different stages. The discriminators are also conditioned on the report embedding $\mathbf{c}$, which takes $\mathbf{c}$ and an image $x$ as input and predicts whether $x$ is real or synthetic.  



\subsection{View Consistency Network (VCN)}
The VCN takes a frontal-view image and a lateral-view image as inputs and generates a score to measure how consistent these two views of images are. Specifically, the level of consistency means how likely that these two images are from the same patient. The weights of the VCN are trained offline. Given a frontal-view image and a lateral-view image in the training data, if they belong to the same patient, we label them as being consistent. They are labeled as inconsistent if belonging to different patients. Given these labeled pairs, we learn a classification network which takes a pair of frontal- and lateral-view images as inputs and judges whether they are consistent. This network first uses a CNN to encode each image separately. Then the embeddings of these two images are fed into a siamese network~\cite{koch2015siamese} to generate a score which indicates how consistent the two images are. This consistency module is added at each stage of the progressive generation process. During training of the progressive generators, we maximize the consistency scores of generated frontal-view and lateral-view image pairs at each stage. 
A modified ResNet-18 \cite{resnet} $f^{(n)}$ is used to compute  image embeddings at stage $n$ in the progressive generation process. Given the embeddings $f^{(n)}(x_f^{(n)})$ and $f^{(n)}(x_l^{(n)})$ of a frontal image $x_f^{(n)}$ and a lateral image $x_l^{(n)}$, the probability that these two images are from the same patient is calculated as: $\sigma(\sum_j w_j|f_j^{(n)}(x^{(n)}_f)-f_j^{(n)}(x^{(n)}_l)|)$,  
where $\sigma(\cdot)$ is the sigmoid function, $j$ indexes the dimensions of embeddings, and $\{w_j\}$ are learnable weights. 

\subsection{Objective Function}
The objective function used for training HAE and MSCGAN consists of three terms: adversarial loss, view-consistency reward, and image reconstruction loss. Note that the view consistency network is trained offline and its weights are frozen during XRayGAN training. For the adversarial loss, we use the loss function in  WGAN-GP~\cite{gulrajani2017improved}, which measures the discrepancy of real images and generated images. The generators aim to minimize this loss and the discriminators aim to maximize this loss. 
The reconstruction loss is defined as the pixel-level L2 distance between a generated image and the groundtruth image. The generators aim to minimize this loss. The view-consistency reward is defined as the probability (in Section 2.3) measuring how likely two images are from the same patient. 
The generators aim to maximize this probability for every pair of generated frontal-view and lateral-view images. A weighted sum of these three terms forms the final objective function. 

\section{Experiments}
\subsection{Datasets}

We evaluate our model on the Open-i dataset~\cite{demner2016preparing} and the  MIMIC-CXR~\cite{DBLP:journals/corr/abs-1901-07042} dataset. Open-i~\cite{demner2016preparing} provides 7,470 chest X-rays with 3,955 radiology reports. We select patient cases who have both a frontal-view X-ray, a lateral-view X-ray, and a report, ending up with 2,585 such cases. 
MIMIC-CXR~\cite{DBLP:journals/corr/abs-1901-07042} contains 377,110 chest X-rays associated with 227,827 radiology reports, divided into subsets. We perform the experiments on the \textit{p10} subset containing 6,654 data examples, each with a frontal-view image, a lateral-view image, and a report. For each dataset, we split it into train, validation, test sets with a radio of 0.7, 0.1, and 0.2. 


\subsection{Experimental Settings} 
In the multi-scale conditional GAN (MSCGAN), the number of stages is set to 4. The image resolution in basic generators is $32\times32$. The resolution of the final image produced by MSCGAN is $256\times256$. In the hierarchical attentional encoder, the size of hidden states in both word-level and sentence-level LSTMs is set to 128. 
To train the networks, we adopt the Adam~\cite{adam} optimizer with  $\beta_1 = 0.9$ and $\beta_2 = 0.999$. The initial learning rates for networks in different stages of MSCGAN are set to $\num{3e-4}, \num{3e-4}, \num{2e-4}$ and $\num{1e-4}$
and reduced by multiplying with 0.2 every 20 epochs. The minibatch sizes for the 4 stages are set to 96, 56, 24, and 12 respectively. The VCNs are trained for 20 epochs with a learning rate of 0.01. In the objective function, we set the weights associated with the adversarial loss, reconstruction loss, and view-consistency reward as 1, 100, and -1 respectively. The model takes 1 days to train with four GeForce GTX 1080 Ti GPUs. Please refer to the supplements for additional hyperparameter settings. 

We compare with three state-of-the-art methods for text-to-image generation: GAN-INT-CLS~\cite{10.5555/3045390.3045503}, StackGAN~\cite{zhang2017stackgan} and AttnGAN~\cite{xu2018attngan}. To ensure a fair comparison, the encoders of texts in these  baselines are set to the hierarchical attentional encoder used in our approach. The generator of GAN-INT-CLS is the same as our basic generator with  8 Res-Upsample  blocks. The StackGAN and AttnGAN models are modified from the official implementation, with architecture unchanged. 


We use four metrics to measure the quality and diversity of generated X-rays, including Inception Score (IS)~\cite{salimans2016improved}, Frechet Inception Distance (FID), Structural Similarity Index (SSIM) \cite{wang2004image}, and View Consistency (VC). VC measures how likely a frontal and a lateral image are from the same patient. To measure VC, we trained a view-consistency network on another subset of MIMIC-CXR.

\begin{table}[t]
\begin{center}
\caption{Results achieved by different methods on the test sets of Open-i and MIMIC-CXR.}
\begin{tabular}{lcccccccc}
\toprule
&\multicolumn{4}{c}{Open-i}&\multicolumn{4}{c}{MIMIC-CXR}\\
\cmidrule(r){2-5}\cmidrule(r){6-9}
Method & IS$\uparrow$ & FID$\downarrow$ & SSIM$\uparrow$& VC $\uparrow$& IS$\uparrow$ & FID$\downarrow$ & SSIM$\uparrow$& VC $\uparrow$\\
\midrule
Real & - & - & - & 0.669 & -&-&-&0.589\\
\midrule
GAN-INT-CLS \cite{10.5555/3045390.3045503} & 1.015&272.7&0.201&0.525 & 1.039&214.3&0.291 & 0.555\\
StackGAN \cite{zhang2017stackgan} & 1.043 &243.4&0.138 &0.487 &1.063&245.5&0.212 &0.471\\
AttnGAN \cite{xu2018attngan} & 1.055 &226.6& 0.171& 0.508 &1.067&232.7&0.231 & 0.487\\
XRayGAN & \textbf{1.081} &\textbf{141.5}&\textbf{0.343}& \textbf{0.627} &\textbf{1.112}&\textbf{86.15}&\textbf{0.379}& \textbf{0.580}\\
\bottomrule
\end{tabular}
\label{table:real_result}
\end{center}
\vspace{-0.4cm}
\end{table}

\subsection{Results}

\begin{wraptable}{r}{0.55\textwidth}
\vspace{-4mm}
\caption{Ablation study on the Open-i dataset.}
\label{table:Ablation Study_result}
    \begin{tabular}{lcccc}
    \toprule
    &\multicolumn{4}{c}{Open-i}\\
    \cmidrule(r){2-5}
    Method & IS$\uparrow$ & FID$\downarrow$ & SSIM$\uparrow$ & VC$\uparrow$ \\
    \midrule
    Real &-&-&-&0.669\\
    \midrule
    w/o VCN & 1.079&147.4& \textbf{0.353}& 0.623\\
    w/o MSCGAN & 1.022 & 235.5 & 0.164 & 0.605\\
    w/o HAE & 1.079 & 153.2 & 0.300& 0.621\\
    \midrule
    Full & \textbf{1.081}& \textbf{141.5}& 0.343 &\textbf{0.627}\\
    \bottomrule
    \end{tabular}
\vspace{-2mm}
\label{table:ablation}
\vspace{-2mm}
\end{wraptable}
Table~\ref{table:real_result} shows the IS (the higher, the better), FID (the lower, the better), SSIM (the higher, the better), and VC (the higher, the better) achieved by different methods on the test sets of Open-i and MIMIC-CXR. From this table, we make the following observations. First, our method achieves much better view consistency (VC) scores than the baselines including GAN-INT-CLS, StackGAN, and AttnGAN. The reason is that our method uses the view-consistency network to encourage image generators to generate frontal-view and lateral-view images yielding large view-consistency probabilities while the baselines lack such a mechanism. Note that the VC scores of real images are higher than the generated images, which is not surprising. Second, our method achieves better IS, FID, and SSIM than baselines. One reason is that the multi-scale conditional GAN in our method is good at generating high-resolution images in a progressive way. The other reason is: the view-consistency network in our method imposes a consistency constraint between frontal-view and lateral-view images. Such a constraint can help to avoid generating insensible images.


\subsubsection{Ablation Studies}

We perform ablation studies on the Open-i dataset to investigate the effectiveness of each module in XRayGAN. In each of the following studies, we change one module with other modules intact.

\vspace{-0.1cm}
\paragraph{View-consistency network} To verify the effectiveness of the view-consistency network (VCN), we remove it from the model. As can be seen from Table~\ref{table:ablation}, removing VCN ("w/o VCN") gets a worse VC score. This demonstrates that VCN is effective for improving view consistency. The reason is straightforward: during training of generators, the VCN explicitly encourages the generators to generate images yielding large VC scores.

\vspace{-0.1cm}

\paragraph{MSCGAN} To verify the effectiveness of MSCGAN, we replace it with a vanilla conditional GAN with the same number of layers and parameters, which generates images at one shot without going through the progressive procedure.  As can be seen from Table~\ref{table:ablation}, this replacement ("w/o MACGAN") leads to substantial degradation of performance on all metrics. This demonstrates that MSCGAN is essential in generating high-resolution  and high-fidelity images. MSCGAN progressively generates a cascade of images with increasing resolutions. This curriculum strategy is better at generating high-resolution images than the at-one-shot generation strategy. 

\vspace{-0.1cm}
\paragraph{Hierarchical LSTM} To verify the effectiveness of hierarchical LSTM, we replace it with a single-layer bidirectional LSTM which treats the entire report as a flat sequence of tokens and ignores the report-sentence-token hierarchical structure. As can be seen from Table~\ref{table:ablation}, this replacement ("w/o HAE") yields worse performance. This demonstrates that using  hierarchical LSTM to capture the linguistic hierarchy in the report can better understand the report, which further makes the generated images have better fidelity. 






\begin{figure}[t]
    \centering
    \vspace{-10mm}
    \subfigure{\includegraphics[width=\linewidth]{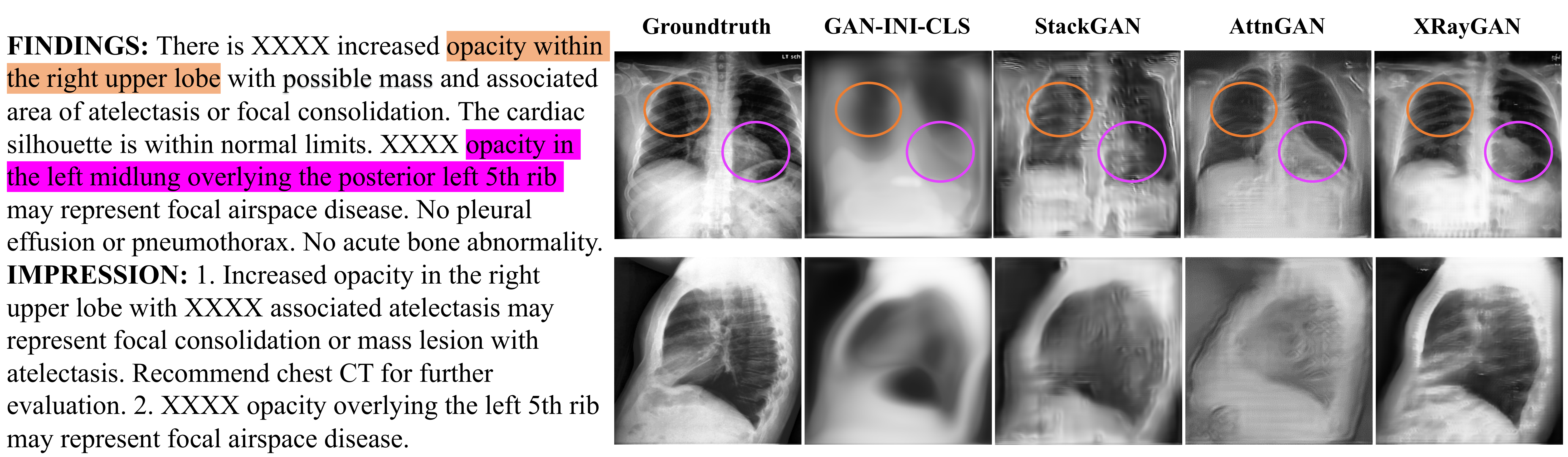}}
    \subfigure{\includegraphics[width=\linewidth]{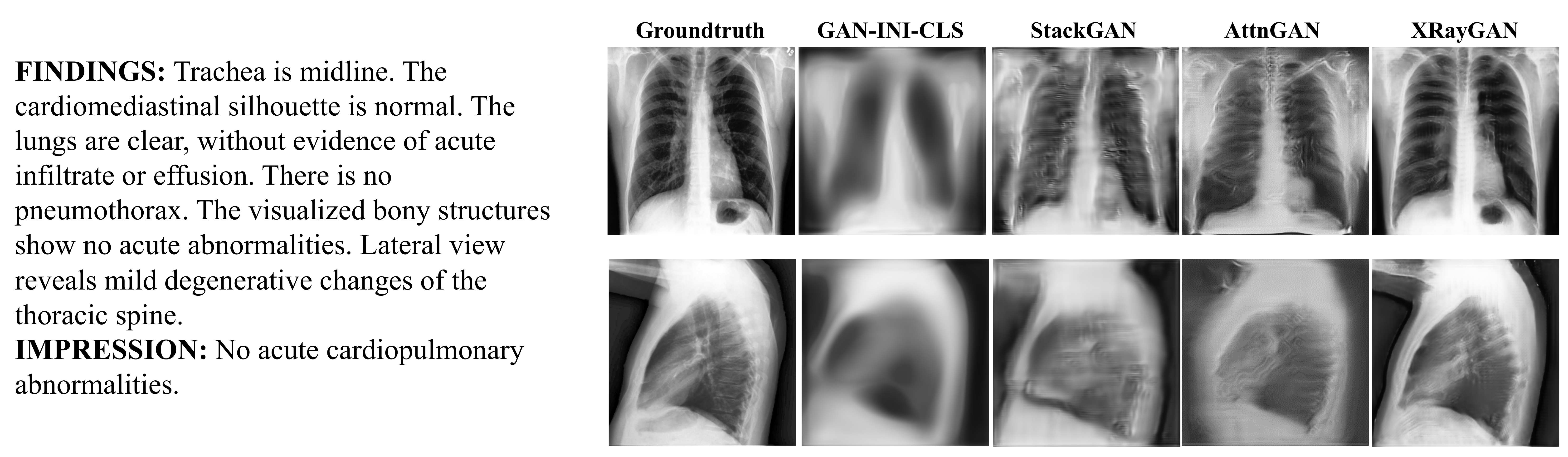}}
    \subfigure{\includegraphics[width=\linewidth]{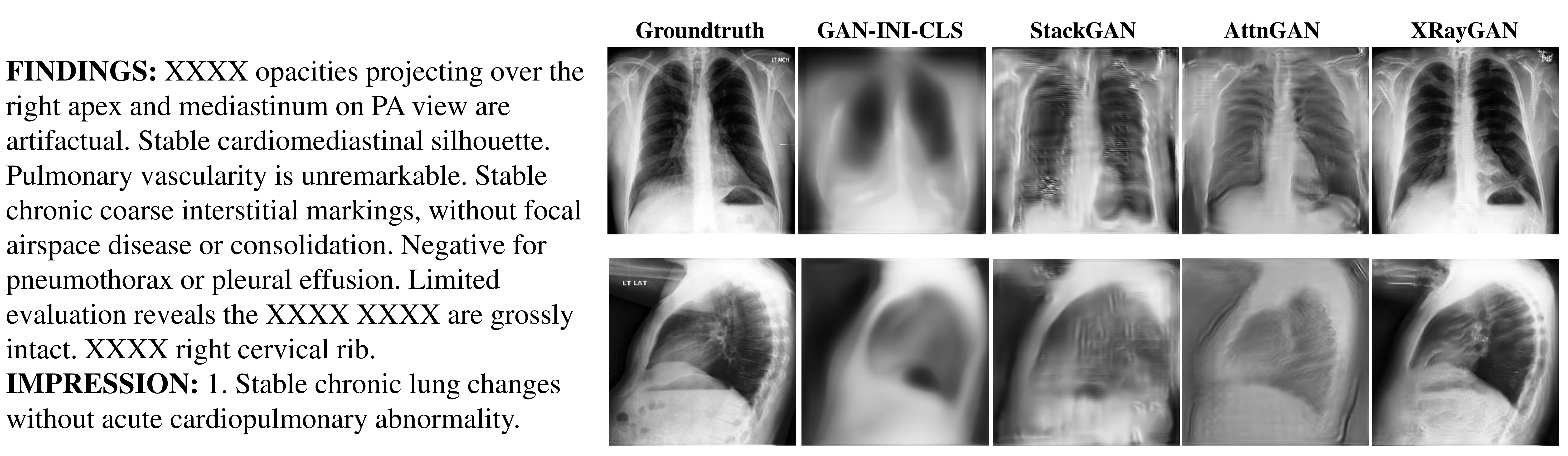}}
    \vspace{-3mm}
    \caption{Frontal-view and lateral-view images generated by different methods from three reports.}
    \label{fig:real visualiztion}
    \vspace{-2mm}
\end{figure}
\subsubsection{Qualitative Evaluation}
 
Figure~\ref{fig:real visualiztion} shows frontal-view and lateral-view images generated by different methods from three reports. From the results, we make the following observations. First, the images generated by our method are clear and visually very similar to the groundtruth, and are more realistic than those generated by the baselines. In particular, our method performs much better in generating lateral-view images than baselines. The lateral-view images generated by baselines have a lot of blur, especially in the lung regions. In contrast, our method successfully generates lateral-view images with clear lungs and ribs. One possible reason that our method works better is: XRayGAN uses a view-consistency network (VCN) to ensure the frontal and lateral images are consistent. If the frontal image is clear, then VCN encourages the lateral image to be clear as well to achieve consistency. The baseline methods lack such a mechanism. On frontal-view images, our method outperforms baselines as well. The frontal-view images generated by GAN-INT-CLS are blurred, where the outline and shadow of  chests can be barely recognized, while those generated by our method are visually clear. In the frontal-view images generated by StackGAN, the outlines of ribs are distorted. In contrast, XRayGAN preserves rib outlines well. In the frontal-view images generated by AttnGAN, the number and morphology of ribs is very different from that in the groundtruth. The ribs generated by our method are visually similar to the groundtruth. The reason that our method generates better frontal-view images is that it uses multi-scale progressive GANs to improve the resolution of generated images. The second observation is that the images generated by XRayGAN correctly reflect the clinical findings in the reports. For example, in the first example, there are two abnormal findings: ``increased opacity within the right upper lobe" and ``opacity in the left midlung overlying the posterior
left 5th rib". The frontal-view image generated by XRayGAN manifests such findings, where the image regions marked by orange and purple circles show the first and second finding respectively. In contrast, images generated by baselines fail to reflect these abnormalities. 

\vspace{-0.1cm}
\section{Related works}
\vspace{-0.1cm}

\vspace{-0.1cm}
\paragraph{Text-to-image generation} Generating images conditioned on texts have been studied in several works~\cite{pmlr-v48-reed16,reed2016learning}. Mansimov et al.~\cite{Mansimov2015GeneratingIF} proposed an encoder-decoder architecture for text-to-image generation. The encoder of text and the decoder of image are both based on recurrent networks. Attention is used between image patches and words. StackGAN~\cite{zhang2017stackgan} first uses a GAN to generate low-resolution images, which are then fed into another GAN to generate high-resolution images. AttnGAN~\cite{xu2018attngan} synthesizes fine-grained details at different subregions of the image by paying attention to the relevant words in the natural language description. DM-GAN~\cite{zhu2019dm} uses a dynamic memory module to refine fuzzy image contents, when the initial images are not well generated and designs a memory writing gate to select the important text information. Obj-GAN~\cite{li2019object} proposes an object-driven attentive image generator to synthesize salient objects by paying attention to the most relevant words in the text description and the pre-generated semantic layout. MirrorGAN~\cite{qiao2019mirrorgan} uses an autoencoder architecture, which generates an image from a text, then reconstructs the text from the image. Different from general-domain text-to-image generation, generating X-rays from radiology reports present unique challenges such as ensuring view consistency, handling the linguistic structure of radiology reports, etc.
\vspace{-0.2cm}
\paragraph{X-ray synthesis} Several attempts have been made to synthesize  X-ray images. In~\cite{madani2018chest,salehinejad2018generalization}, GANs are used to generate X-ray images from random vectors. Ganesan et al.~\cite{ganesan2019assessment} uses progressively-growing GAN to synthesize high-resolution X-rays. Malygina et al.~\cite{10.1007/978-3-030-37334-4_29} uses CycleGAN to augment X-rays images for infrequent diseases. Ying~\cite{Ying_2019_CVPR} reconstructs CTs from biplanar X-rays using GANs. Our work differs from these works in two aspects: (1) we generate X-rays from radiology reports; (2) we generate both frontal-view and lateral-view images.
\vspace{-0.2cm}
\paragraph{Image generation} GANs~\cite{goodfellow2014generative} have been widely used for image generation from random vectors. Conditional GANs~\cite{Mirza2014ConditionalGA,odena2017conditional} generates images from class labels. Image-to-image translation~\cite{pix2pix2016,wang2018pix2pixHD,CycleGAN2017} studies how to generate one image (set) from another (set) based on GANs. Brock et al.~\cite{brock2018large} demonstrate that GANs benefit dramatically from scaling: increasing model size and minibatch size improves the fidelity of generated images.



\vspace{-0.2cm}
\section{Conclusions}
\vspace{-0.3cm}
In this paper, we study how to generate view-consistent, high-resolution, and high fidelity X-ray images from radiology reports, for the sake of facilitating training of medical students specialized in radiology. We propose an XRayGAN where a view consistency network is used to encourage the generated frontal-view and lateral-view images to be from the same patient, a multi-scale conditional GAN is used to generate high-resolution X-ray images in a progressive manner, and a hierarchical attentional encoder is used to capture the hierarchical linguistic structure and various clinical importance of words and sentences. We demonstrate the effectiveness of our methods on two radiology datasets.

\bibliographystyle{unsrt}
\bibliography{refs}

\end{document}